\documentclass{elsart}
\usepackage{graphicx,amssymb}
\journal{Astroparticle Physics}

\begin{document} 

\begin{frontmatter}
\title {Supernova Neutrino Detection in Borexino}
\author{L. Cadonati\thanksref{Cadonati}},
\author{F.P. Calaprice}, 
\author{M.C. Chen\thanksref{Chen}}
\address{Department of Physics, Jadwin Hall, 
         Princeton University, Princeton, NJ 08544}
\thanks[Cadonati]
     {Corresponding author:  
         Department of Physics, Jadwin Hall, 
         Princeton University, Princeton, NJ 08544, USA;
         telephone 1-609-258-1122; fax 1-609-258-1356
         email cadonati@princeton.edu}
\thanks[Chen]
     {Present address: Department of Physics, Stirling Hall,
      Queen's University, Kingston, Ontario, Canada.}

\begin{abstract}
We calculated the expected neutrino signal in Borexino from a 
typical Type II supernova at a distance of $10\,$kpc. 
A burst of around 110 events would appear in Borexino within 
a time interval of about $10\,$s. 
Most of these events would come from the reaction channel 
$\bar{\nu}_e+p\rightarrow e^++n$, while about 30 events would 
be induced by the interaction of the supernova neutrino flux 
on $^{12}$C in the liquid scintillator. 
Borexino can clearly distinguish between the neutral-current 
excitations 
$^{12}\mbox{C}(\nu,\nu')^{12}\mbox{C}^*\,(15.11\,{\rm MeV})$ 
and the charged-current reactions 
$^{12}\mbox{C}(\nu_e,e^-)^{12}\mbox{N}$ and 
$^{12}\mbox{C}(\bar{\nu}_e,e^+)^{12}\mbox{B}$, via their 
distinctive event signatures. 
The ratio of the charged-current to neutral-current neutrino 
event rates and their time profiles with respect to each other 
can provide a handle on supernova and non-standard neutrino 
physics (mass and flavor oscillations).
\end{abstract}

\begin{keyword}
Neutrino detector \sep Supernovae \sep Neutrino mass
\PACS 14.60.Pq \sep 25.30.Pt \sep 95.55.Vj \sep 97.60.Bw
\end{keyword}
\end{frontmatter}

\newpage
\section{Introduction}

The last supernova in our galaxy that was visible on Earth was
observed in 1604.
Back then, astronomers marveled at ``Kepler's Star'' in Ophiuchus 
and studied the light curve from the colossal explosion.  
The next Galactic supernova will differ from previous ones in 
human history.  
Not only will astronomical observations be made across the 
electromagnetic spectrum; it is likely that several neutrino 
detectors will detect the burst of neutrinos emerging from the 
stellar gravitational collapse.  
Indeed, it may turn out that the next Galactic supernova will 
not be seen optically, due to obscuration (as was probably the 
case for the radio source supernova remnant Cassiopeia A), and 
that supernova neutrino detection will signal the astrophysical 
event, triggering its search in electromagnetic bands.

The field of extrasolar neutrino astrophysics was born when 
neutrinos from SN1987A in the Large Magellanic Cloud were 
detected by the IMB~\cite{IMB} and Kamiokande~\cite{Kamioka} 
neutrino detectors.  
These pioneering observations contributed significantly to our 
understanding of the mechanisms involved in a supernova 
explosion, as well as providing interesting limits on neutrino 
properties.  
The next Galactic supernova will prove even more valuable owing 
to the abundance of neutrino events produced by the closer source 
and the variety of reactions that will be available to study 
these neutrinos.

SuperKamiokande~\cite{SuperK} and SNO~\cite{SNO} will be the 
largest supernova neutrino detectors in the foreseeable future.  
SuperKamiokande contains about 50,000 tons of water, providing 
proton targets for the inverse $\beta$ decay reaction.
The SNO detector is filled with a kiloton of heavy water that 
offers deuterons for neutrino interactions.
Detectors employing liquid scintillator that will be capable of 
detecting supernova neutrinos include Baksan~\cite{Baksan},
LVD~\cite{LVD}, MACRO~\cite{MACRO}, 
Borexino~\cite{Borexino}, and KamLAND~\cite{KamLAND}.  
In a liquid scintillator (typical composition CH$_2$),
the antineutrino-proton reactions will still constitute the 
majority of the detected supernova neutrino events.  
Nevertheless, the abundance of carbon in an organic liquid 
scintillator provides an additional, interesting target for 
neutrino interactions~\cite{japan}.

Neutrino reactions on the $^{12}$C nucleus 
${\rm (J^\pi=0^+,T=0)}$ include the superallowed transitions to 
the A=12 triad ${\rm (J^\pi=1^+,T=1)}$ of $^{12}$B 
(ground state), $^{12}\mbox{C}^*$ ($15.1\,$MeV) and $^{12}$N 
(ground state).  
The cross sections for these reactions are reasonably large, 
for neutrinos above the reaction thresholds of 14.4, 15.1, and 
$17.3\,$MeV, respectively.  
This system is in many ways ideal for investigating neutrino 
physics since it offers charged-current and neutral-current 
neutrino reactions, allowing one to probe for potential neutrino 
flavor oscillations.
In addition, each of the above reactions has a distinguishing
detection signature.  
In the case of the charged-current reactions, the signature is 
the electron (or positron) emitted in the prompt reaction, 
followed  tens of milliseconds later by the $\beta^+$ 
decay (or $\beta^-$ decay) back to $^{12}$C --- a delayed 
coincidence that enables identification of these specific 
interactions from background.  For the neutral-current 
excitation of $^{12}$C, the $15.1\,$MeV gamma following 
de-excitation is the signature.

In a Type II supernova, neutrinos are emitted with a thermal 
spectrum.  
The average neutrino energy depends upon the species of 
neutrino, since decoupling in the neutrinosphere occurs at 
different temperatures for the different species.
Typical values for the average neutrino energies 
are~\cite{langvog}: 
$\langle E \rangle = 11\,$MeV for $\nu_e$, 
$\langle E \rangle = 16\,$MeV for $\bar{\nu}_e$, and 
$\langle E \rangle = 25\,$MeV for $\nu_\mu$, $\nu_\tau$, 
$\bar{\nu}_{\mu}$, and $\bar{\nu}_{\tau}$.  
This energy hierarchy enhances the sensitivity of the 
neutrino-carbon reactions to non-standard neutrino physics. 
The $\nu_\mu$'s and $\nu_\tau$'s are more energetic than
electron neutrinos. 
They dominate the neutral-current reaction
$^{12}\mbox{C}(\nu,\nu')^{12}\mbox{C}\,(15.11\,{\rm MeV})$
with an estimated contribution of around 90\%.
This allows the $^{12}$C $15.1\,$MeV events to be closely 
identified with the ``heavy flavor'' species.  
Additionally, since the average energy of the $\nu_e$ or 
$\bar{\nu}_e$ from the supernova is at or below their 
charged-current reaction thresholds, the total number of 
these events will be small.  
This provides a pseudo-appearance opportunity for observing 
neutrino flavor oscillations.

In order to exploit these aspects for exploring non-standard
neutrino physics, a liquid scintillator supernova neutrino 
detector needs to be able to cleanly detect the 
$15.1\,$MeV $\gamma$ ray from the neutral-current reaction.
This implies that the detector requires a large volume to 
contain this energetic $\gamma$ ray.  
Additionally, good energy resolution is necessary 
for resolution of the peak produced by these events in the 
energy spectrum.
The LVD and MACRO experiments are existing scintillator 
detectors, but their $15.1\,$MeV $\gamma$ ray capabilities
are somewhat limited since they are both segmented detectors.
Borexino and KamLAND will have excellent efficiency for 
detecting and resolving the $15.1\,$MeV signal, as they will 
both feature large, homogeneous volumes of liquid scintillator.

In this paper, we present calculations of the neutrino 
event rates in Borexino for a hypothetical Type II supernova 
at a distance of $10\,$kpc, with a typical binding energy release 
of $3\times 10^{53}\,$ergs.  
We illustrate the neutral-current capability of Borexino, via 
detection of the $15.1\,$MeV $\gamma$ ray.  
Finally, we describe the capabilities of Borexino for 
exploring supernova neutrino oscillations and setting 
stringent neutrino mass limits, based upon the delayed 
arrival of a heavy $\nu_\tau$.  
Despite being a much smaller detector, the neutrino mass 
limits to which Borexino is sensitive are comparable to what 
might be achieved by Super-Kamiokande and SNO, in the event of 
a Galactic supernova.

\section{Supernova Neutrino Signatures in Borexino}

The Borexino experiment~\cite{Borexino} is under
construction at the Laboratori Nazionali del Gran Sasso
underground facility, in Italy.
Its main physics goal is the detection of the 0.86 MeV $\nu_e$ 
from electron-capture decay of $^7$Be in the Sun.
The detector has a central volume containing $300\,$tons of
liquid scintillator (pseudocumene plus fluor), contained in a
spherical nylon bag and viewed by 2,200 photomultiplier tubes. 

Supernova neutrinos will interact in the liquid scintillator via
electron scattering, inverse $\beta$ decay of the proton and 
reactions on $^{12}$C\@.  
Constructed as a low-background solar neutrino detector,
Borexino will be able to cleanly detect all of these supernova
neutrino interactions, with the exception that the handful of 
electron scattering events have no clear distinguishing 
signature.

We start by considering a stellar gravitational collapse 
releasing $\varepsilon_B\approx 3\times 10^{53}\,$ergs binding 
energy~\cite{comm}.
$99\%$ of this binding energy comes off in the form of 
neutrinos, most of which originate from $\nu\bar{\nu}$ pair 
production.
Only a small fraction (a few percent) of the neutrinos are 
produced in other processes, such as the neutronization during 
core collapse.

At core collapse temperatures and densities exceeding 
$10^{11}\,{\rm g/cm^3}$, matter is not transparent to neutrinos.  
Scattering interactions thermalize the neutrinos though their 
mean free path remains large; in effect, neutrinos become the 
energy transport agents in the collapsed stellar core.  
They emerge from the cooling core or ``neutrinosphere'' after 
they decouple.  The temperature at the time of decoupling 
determines the energy distribution of the emitted neutrinos.
Neutrinos of $\mu$ and $\tau$ flavor decouple at higher 
temperature, since they interact in ordinary matter only via 
the neutral-current weak interaction, whereas charged-current 
scattering can occur for $\nu_e$ and $\bar{\nu}_e$.  
Moreover, the neutrino decoupling takes place in neutron rich 
matter, which is less transparent to $\nu_e$ than 
$\bar{\nu}_e$.  
The temperature hierarchy is, then: 
${\rm T}_{\nu_e}<{\rm T}_{\bar{\nu}_e}<{\rm T}_{\nu_{x}}$,
where $\nu_x$ denotes $\nu_{\mu,\tau}$ and 
$\bar{\nu}_{\mu,\tau}$.
Each spectrum can be considered as a Fermi-Dirac distribution
with zero chemical potential~\cite{langvog}:
\begin{equation}
\label{fdist}
{\rm \frac{dN}{dE_{\nu}}=\frac{0.5546}{T^3} \frac{E_{\nu}^2}
{1+{\rm e}^{E_{\nu}/T}}\, N_0},
\end{equation}
with the following parameters:
\begin{eqnarray}
\nu_e \;\;\;      & {\rm T=3.5\, MeV}\;\;\;  
         & {\rm \langle{\rm E_{\nu}}\rangle = 11\, MeV}; \\
\bar{\nu}_e\;\;\; & {\rm T=5\, MeV}\;\;\;\;\; 
         & {\rm \langle{\rm E_{\nu}}\rangle = 16\, MeV};\\
\nu_{\mu ,\tau}\;\bar{\nu}_{\mu ,\tau}\;\;\; 
         & {\rm T=8\, MeV}\;\;\;\;\; & 
         {\rm \langle{\rm E_{\nu}}\rangle = 25\, MeV.} 
\end{eqnarray}

Based upon equipartition, the prediction is that all of the
neutrino species are produced in the cooling core with the
same luminosity~\cite{comm}.  This implies that the number of
$\nu_e$'s will be greater than $\nu_{\mu}$ and $\nu_{\tau}$,
since their average energy is lower.

The expected event rates for each of the reactions in Borexino 
were calculated by integrating the cross sections as functions 
of energy over the Fermi-Dirac spectra for the relevant neutrino 
species.
The total number of neutrinos of each type was estimated by 
dividing the binding energy equally between the six $\nu$ or 
$\bar{\nu}$ species, and then dividing by the average energy 
${\rm \langle E\rangle}$ of each species.
We consider $300\,$tons of pseudocumene (C$_9$H$_{12}$)
as the target for neutrinos in Borexino and list the reactions,
thresholds, and number of targets in Table~\ref{targets}.

\begin{table}
\caption {Supernova neutrino reactions in Borexino, their 
energy thresholds and the number of targets in the sensitive 
volume ($300\,$tons C$_9$H$_{12}$).}
\label{targets}
\begin{tabular}{lcc}
\hline
Reaction & E$_{thres}$ & Number of targets in $300\,$tons \\
\hline
$\nu + e^- \rightarrow \nu + e^-$        &  0       
                                & $9.94\times 10^{31}$ \\
$\bar{\nu}_e + p \rightarrow e^+ + n$    & $1.80\,$MeV 
                                 & $1.81\times 10^{31}$\\
$^{12}\mbox{C}(\nu_e,e^-)^{12}\mbox{N}$  & $17.3\,$MeV 
                                 & $1.36\times 10^{31}$ \\
$^{12}\mbox{C}(\bar{\nu}_e,e^+)^{12}\mbox{B}$  & $14.4\,$MeV                                        
                                 & $1.36\times 10^{31}$ \\
$^{12}\mbox{C}(\nu,\nu')^{12}\mbox{C}^*$       & $15.1\,$MeV 
                                 & $1.36\times 10^{31}$ \\ 
\hline
\end{tabular}
\end{table}


\subsection{$\nu - e^-$ scattering}
\label{nuesc}

Neutrino-electron scattering produces recoil electrons with 
energy from zero up to the kinematic maximum.  
In Borexino the recoil electron detection threshold will be 
$0.25\,$MeV, small compared to supernova neutrino energies.
In our rate calculation we approximate by integrating over all 
electron recoil energies.  
The standard electroweak cross section, with
${\rm E_\nu \gg m}_e$, is: 
\begin{equation}
\sigma = \frac{2 \, {\rm G_F^2 m}_e {\rm E}_{\nu}}{\pi} 
{\rm \left[ c_L^2 + \frac{1}{3} c_R^2 \right],}
\end{equation}
where the coupling constants depend on the neutrino species
considered.  The total cross sections for $\nu-e^-$ scattering
are linearly proportional to the neutrino energy, and appear
with appropriate parameters as: 
\begin{eqnarray}
\sigma (\nu_e e \to \nu_e e) &=&
       9.20\times 10^{-45} \, {\rm E_{\nu\,[MeV]}\,cm^2; }\\
\sigma (\bar{\nu}_e e \to \bar{\nu}_e  e) &=&
       3.83\times 10^{-45} \, {\rm E_{\nu\,[MeV]}\,cm^2; }\\ 
\sigma (\nu_{\mu ,\tau} e \to \nu_{\mu ,\tau} e) &=&
       1.57\times 10^{-45} \, {\rm E_{\nu\,[MeV]}\,cm^2; }\\
\sigma (\bar{\nu}_{\mu ,\tau} e \to \bar{\nu}_{\mu ,\tau} e) &=&
       1.29\times 10^{-45} \, {\rm E_{\nu\,[MeV]}\,cm^2. }
\end{eqnarray}
For all of the neutrino species from a typical supernova at 
$10\,$kpc, the calculated event rate in Borexino is about 
5 events from neutrino-electron scattering.


\subsection{$\bar{\nu}_e + p$ reaction}
\label{nupsc}

The large cross section, low threshold, and abundance of target 
protons makes this the dominant channel for detection of 
supernova neutrinos.
The inverse $\beta$ decay of the proton:
\begin{equation}
\bar{\nu}_e + p \rightarrow e^+ + n,
\end{equation}
has a reaction threshold of ${\rm E_{thres}=1.80}\,$MeV.
At low energies we approximate the total cross section as:
\begin{equation} \label{nuecs}
\sigma = \frac{{\rm G_F^2 p}_e {\rm E}_e}{\pi}
\left| \cos^2{\theta_c} \right|^2
\left[ 1+3 \left( \frac{\rm g_A}{\rm g_V} \right)^2\right],
\end{equation}
which can be re-written with appropriate parameters as:
\begin{equation}
{\rm \sigma(E_{\nu})=
9.5\times 10^{-44}(E_{\nu\,[MeV]}-1.29)^2\; cm^2.}
\end{equation}
Integrating this cross section with the $\bar{\nu}_e$
spectrum gives an event rate of about 79 neutrinos in Borexino 
from a typical supernova burst.

\subsection{$^{12}\mbox{C}$ reactions}\label{cscat}
   
\begin{figure}
\includegraphics[width=4in]{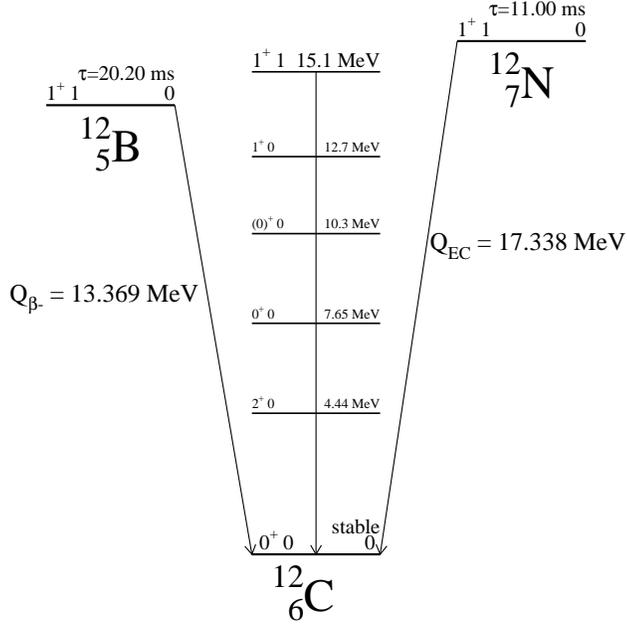}
\caption
{Level diagram for the $^{12}$C, $^{12}$N, $^{12}$B triad.}
\label{isobars}
\end{figure}

The A=12 isobar level scheme is shown in Fig.~\ref{isobars}.
Superallowed transitions leading from the $^{12}\mbox{C}(0^+,0)$ 
ground state to the triad of $(1^+,1)$ states: 
${\rm ^{12}N_{g.s.}}$, ${\rm ^{12}B_{g.s.}}$ and 
${\rm ^{12}C^*(15.11\,MeV)}$ give rise to the following three 
neutrino reactions:
\begin{enumerate}
\item[1)] charged-current capture of $\bar{\nu}_e$:
\begin{eqnarray}
\bar{\nu}_e +^{12}\mbox{C} \to ^{12}\mbox{B} + e^+ 
    &\;\;\;\;\;\;\;{\rm Q=14.39\,MeV}\\
^{12}\mbox{B} \to ^{12}\mbox{C}+e^-+\bar{\nu}_e  
    &\;\;\;\;\;\;\;\tau_{1/2}=20.20\,{\rm ms};
\end{eqnarray}
\item[2)] charged-current capture of $\nu_e$:
\begin{eqnarray}
\nu_e +^{12}\mbox{C}\to ^{12}\mbox{N} + e^- 
    &\;\;\;\;\;\;\;{\rm Q=17.34\, MeV}\\
^{12}\mbox{N} \to ^{12}\mbox{C}+e^++\nu_e 
    &\;\;\;\;\;\;\;\tau_{1/2}=11.00\,{\rm ms};
\end{eqnarray}
\item[3)] neutral-current inelastic scattering of $\nu$ or 
$\bar{\nu}$:
\begin{eqnarray}
\nu +^{12}\mbox{C} \to ^{12}\mbox{C}^* + \nu' 
    &\;\;\;\;\;\;{\rm Q=15.11\, MeV}\\
^{12}\mbox{C}^* \to ^{12}\mbox{C} + \gamma
    &\;\;\;(15.11\, {\rm MeV)}.
\end{eqnarray}
\end{enumerate}

All of the reactions on carbon can be tagged in Borexino.  The
charged-current events have the delayed coincidence of a
$\beta$ decay following the interaction.  
The neutral-current events have a monoenergetic $\gamma$ ray at 
$15.1\,$MeV.

In the low-energy limit (${\rm E\ll 250\, MeV}$)
only superallowed and allowed transitions are significant;
the charged-current cross section is given by:
\begin{equation}\label{chargedcs} 
\sigma = 
\frac {\rm G_F^2}{\pi}cos^2\theta_c {\rm \sum_{i}|M_i|^2}
{\rm p}_e {\rm E}_e {\rm F(Z,E}_e)
\end{equation}
where ${\rm |M_i|^2}$ are the nuclear matrix elements squared
and ${\rm F(Z,E}_e)$ is the Fermi function, accounting for Coulomb 
corrections in $\beta$ decays.
For the neutral-current reaction only the isovector axial 
current contributes to the interaction; the cross section is 
given by:
\begin{equation}\label{neutralcs}{\rm
\sigma=\frac{G_F^2}{\pi} \sum_i |M_i|^2 (E_{\nu}-E_i)^2}.
\end{equation}

The cross sections for the neutrino-carbon reactions have been
investigated theoretically and experimentally over the past
20 years and are now well established.  
The nuclear matrix element is the Gamow-Teller matrix element, 
determined at $q^{2}=0$ by the  $\beta$ decay rates, in the
Elementary Particle Treatment~\cite{japan,theocc}. 
The resulting cross sections are confirmed also by the other two 
theoretical approaches:  the Shell 
Model~\cite{vogel,shell,donnelly,walecka}
and the Random Phase Approximation~\cite{rpa}.

\begin{table}
\caption{Neutrino-carbon cross sections as measured and 
calculated by different authors, in units of $10^{-42}\,$cm$^2$. 
These are averaged over the experimental neutrino spectrum 
originating from muon decay at rest.} 
\label{cs}
\begin{tabular}{lcc}
\hline   
    & $^{12}\mbox{C}(\nu_e,e^-)^{12}\mbox{N}_{\rm g.s.}$ & 
           $^{12}\mbox{C}(\nu,\nu')^{12}\mbox{C}^*
	                    \;[\nu=\nu_e+\bar{\nu}_{\mu}]$ \\
\hline
EPT~\cite{japan,theocc}       
    &  9.2                    &  9.9  \\
Shell model~\cite{vogel} 
    &  9.1                    &  9.8  \\
RPA~\cite{rpa}           
    &  9.3                    &  10.5 \\
KARMEN~\cite{karmen}    
    & $ 8.9 \pm 0.6 \pm 0.75$ & $11 \pm 1.0 \pm 0.9$ \\
LAMPF~\cite{lampf}       
    & $10.5 \pm 1.0 \pm 1.0 $ &           --         \\
LSND~\cite{lsnd}         
    & $ 9.1 \pm 0.4 \pm 0.9 $ &           --         \\
\hline
\end{tabular}
\end{table}

Measurements of cross sections for 
$^{12}\mbox{C}(\nu_e,e^-)^{12}\mbox{N}$
and $^{12}\mbox{C}(\nu,\nu')^{12}\mbox{C}^*$ have also 
been performed at KARMEN~\cite{karmen}, at LAMPF~\cite{lampf},
and by LSND~\cite{lsnd}.  Table~\ref{cs} compares the 
theoretical predictions to the measured data.  
The agreement is good and we based our rate calculations on a 
combined theoretical and experimental average of
${\rm\langle\sigma\rangle_{exp}=9.2\times 10^{-42}\, cm^2}$.
The cross section measurements were averaged over the
neutrino energies relevant to the experiments, namely neutrinos 
from muon decay at rest.  
It is straightforward to scale these measured values
to give averaged cross sections for supernova neutrinos.

Since $^{12}$N and $^{12}$B are ``mirror nuclei'', 
the matrix elements and energy-independent terms in the cross 
section are essentially identical.  
Only the Coulomb correction differs when calculating the 
$\bar{\nu}_e$ capture rates.

The neutral-current cross section can also be extracted from the 
experimental $^{12}\mbox{C}(\nu,\nu')^{12}\mbox{C}^*$ cross 
section data.
Note that the quoted values for $\langle\sigma\rangle$ from
Table~\ref{cs} are averaged over the experimental neutrino flux 
including both $\nu_e$ and $\bar{\nu}_{\mu}$ from muon decay.
Using an averaged value
${\rm \langle\sigma\rangle_{exp}=10\times 10^{-42}\, cm^2}$
and assuming the contribution of $\nu_e$ and $\bar{\nu}_{\mu}$ 
to the nuclear inelastic scattering to be the same, we scaled 
this data for supernova neutrino fluxes and energies.

\begin{table} 
\caption{Supernova neutrino events in Borexino from a
supernova at $10\,$kpc, with
$\varepsilon_B = 3\times 10^{53}\,$ergs binding energy release.}
\label{c12}
\begin{tabular}{cccc}
\hline
reaction channel & $\langle{\rm E_{\nu}}\rangle$ [MeV] 
   &$\langle\sigma\rangle$ [cm$^2$] & ${\rm N_{events}}$ \\
\hline
$\nu_e-e$        & 11 & $ 1.02\times 10^{-43}$ & 2.37 \\
$\bar{\nu}_e-e$  & 16 & $ 6.03\times 10^{-44}$ & 0.97 \\
$\nu_x-e$        & 25 & $ 3.96\times 10^{-44}$ & 0.81 \\
$\bar{\nu}_x-e$  & 25 & $ 3.25\times 10^{-44}$ & 0.67 \\
\hline
total $\nu-e$    &    &                        & 4.82 \\
\hline
                 &    &                        &      \\
$\bar{\nu}_e+p \rightarrow e^+ + n$ 
                 & 16 & $ 2.70\times 10^{-41}$ & 79   \\
                 &    &                        &      \\
\hline
                 &    &                        &      \\
$^{12}\mbox{C}(\nu_e,e^-)^{12}\mbox{N}$       
                  & 11 & $1.85\times 10^{-43}$ & 0.65 \\
$^{12}\mbox{C}(\bar{\nu}_e,e^+)^{12}\mbox{B}$ 
                 & 16 & $1.87\times 10^{-42}$  & 3.8 \\
                 &    &                        &      \\
\hline
neutral-current excitation    &    &        &    \\
$\nu_e + ^{12}\mbox{C}$       
    & 11 & $1.33\times 10^{-43}$ & 0.4 \\
$\bar{\nu}_e+^{12}\mbox{C}$   
    & 16 & $6.88\times 10^{-43}$ & 1.5 \\
$\nu_x+^{12}\mbox{C}$         
    & 25 & $3.73\times 10^{-42}$ & 20.6\\
\hline
total $^{12}\mbox{C}(\nu,\nu')^{12}\mbox{C}^*$ 
    &    &      & 22.5\\
\hline
\end{tabular}
\end{table}

Table~\ref{c12} summarizes our results; we estimate 23 
neutral-current events, 4 events due to $\bar{\nu}_e$
capture on $^{12}\mbox{C}$ and less than one event due to
$\nu_e$ capture, from a typical Galactic supernova at $10\,$kpc.

\section{Neutral Current Detection in Borexino}
   
The neutrino burst from a supernova rises steeply and decays 
exponentially in time: ${\rm L}_{\nu}\sim {\rm e}^{-t/{\tau_{\nu}}}$, 
with $\tau_{\nu}\approx 3$~s~\cite{emtime}.  
In a low-background solar neutrino detector, a burst of 100 
events in a time window of 10 seconds is easily identified.  
The ability to separate the neutral-current events in a liquid 
scintillator detector from the $\bar{\nu}_e - p$ reactions 
determines whether interesting neutrino physics can be explored.

The inverse $\beta$ decay of the proton produces a neutron.
In Borexino, this neutron thermalizes and walks in the detector 
until it is captured by hydrogen: $n+p \to d + \gamma$,
with a mean capture time $\tau = 250\,\mu$s and 
E$_{\gamma}=2.2\,$MeV\@.
The large homogeneous detection volume in Borexino ensures 
efficient neutron capture and efficient detection of the 
$2.2\,$MeV $\gamma$.  
These events can be tagged by the delayed coincidence between 
the initial $e^+$ from the $p(\bar{\nu}_{e},e^+)n$ reaction
and the neutron capture $\gamma$ ray.

On the other hand, if a detector lacks the low-energy threshold 
of Borexino or a detector is not able to contain the neutron 
produced by the $\bar{\nu}_e - p$ reaction, it will not be able 
to exploit the delayed coincidence signature to identify these 
events.
Consequently, the $\bar{\nu}_e - p$ events appear as single 
positrons.
The challenge in such a detector is then to distinguish the
$15.1\,$MeV $\gamma$ of $^{12}\mbox{C}$ neutral-current 
excitation from the continuum spectrum produced by these 
positrons.

\begin{figure}
\includegraphics[width=4.5in]{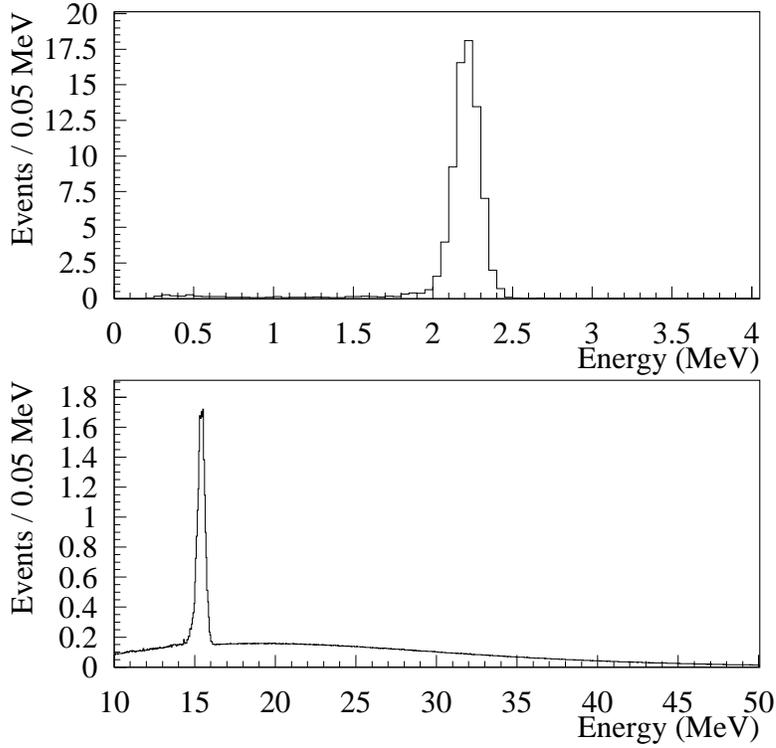}
\caption{Simulated singles spectrum from supernova neutrinos in 
Borexino, in two different energy windows.
In the lower energy spectrum there will be a 
peak at $2.2\,$MeV, due to the capture of neutrons produced 
in the charged current $p(\bar{\nu}_{e},e^+)n$ reaction.
In the higher energy end of the spectrum (E$>10\,$MeV), 
the $15.1\,$MeV $\gamma$ rays from neutral-current excitation of
carbon are well resolved from the continuum of $e^+$ events
from $p(\bar{\nu}_e,e^+)n$.}
\label{spec}
\end{figure}

Efficient detection and resolution of the $15.1\,$MeV 
$\gamma$ will also be possible in Borexino.  
Fig.~\ref{spec} depicts an example of the singles spectrum from 
all supernova neutrino events that would occur
in Borexino within a time window of 10 seconds.
Even if the $\bar{\nu}_e - p$ positrons are
not tagged by the delayed neutron capture $\gamma$ ray, 
the $15.1~$MeV peak is well resolved on top of the
positron spectrum.  The energy resolution in Borexino,
simulated here with the design light collection statistics
of 400 photoelectrons/MeV, allows the neutral-current events
to be identified.  The large, homogeneous volume of liquid
scintillator effectively contains the total energy of this
$\gamma$ ray.

\section{Consequences of Non-Standard Neutrino Physics}
\label{newphys}
\subsection{Neutrino Mass from Time of Flight}
   
The present limits on neutrino mass, obtained by laboratory 
experiments, are high for $\nu_{\mu}$ and $\nu_{\tau}$:
${\rm m}_{\nu_e}<3.9\,{\rm  eV}$, 
${\rm m}_{\nu_{\mu}}< 170\,{\rm  keV} $ and 
${\rm m}_{\nu_{\tau}} < 18.2\,{\rm MeV}$~\cite{pdb}.
By studying the arrival time of neutrinos of different flavors 
from a supernova, mass limits on $\nu_{\mu}$ and $\nu_{\tau}$ 
down to the tens of eV level can be explored.

Consider a supernova neutrino flux composed of two species, one 
massive and the other essentially massless.
The massive neutrinos will reach Earth with a time delay:
\begin{equation}\label{deltat}{\rm
\Delta t = \frac {D}{2c} \left( \frac{m_{\nu}}{E_{\nu}} 
\right)^2
}\end{equation}
with respect to the massless species, where D is the distance 
to the supernova.  
Measuring this time delay requires being able to distinguish the 
massive species from the massless neutrino interactions.  
Ideally, knowledge of the emission time distribution
is also required as is a precise measurement of E$_\nu$.

\begin{figure}
$\begin{array}{c c} 
\includegraphics[width=2.5in]{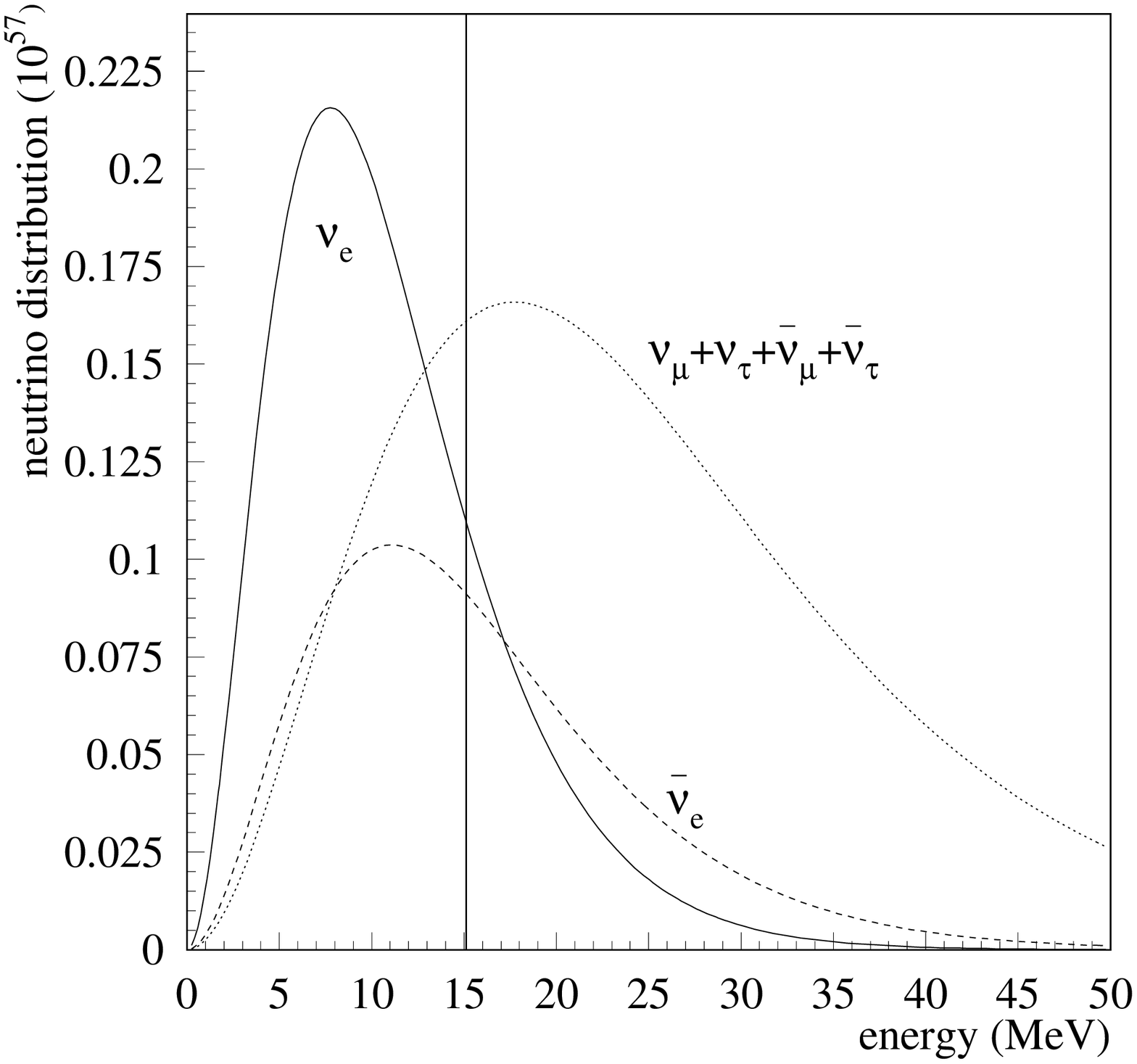} &
\includegraphics[width=2.5in]{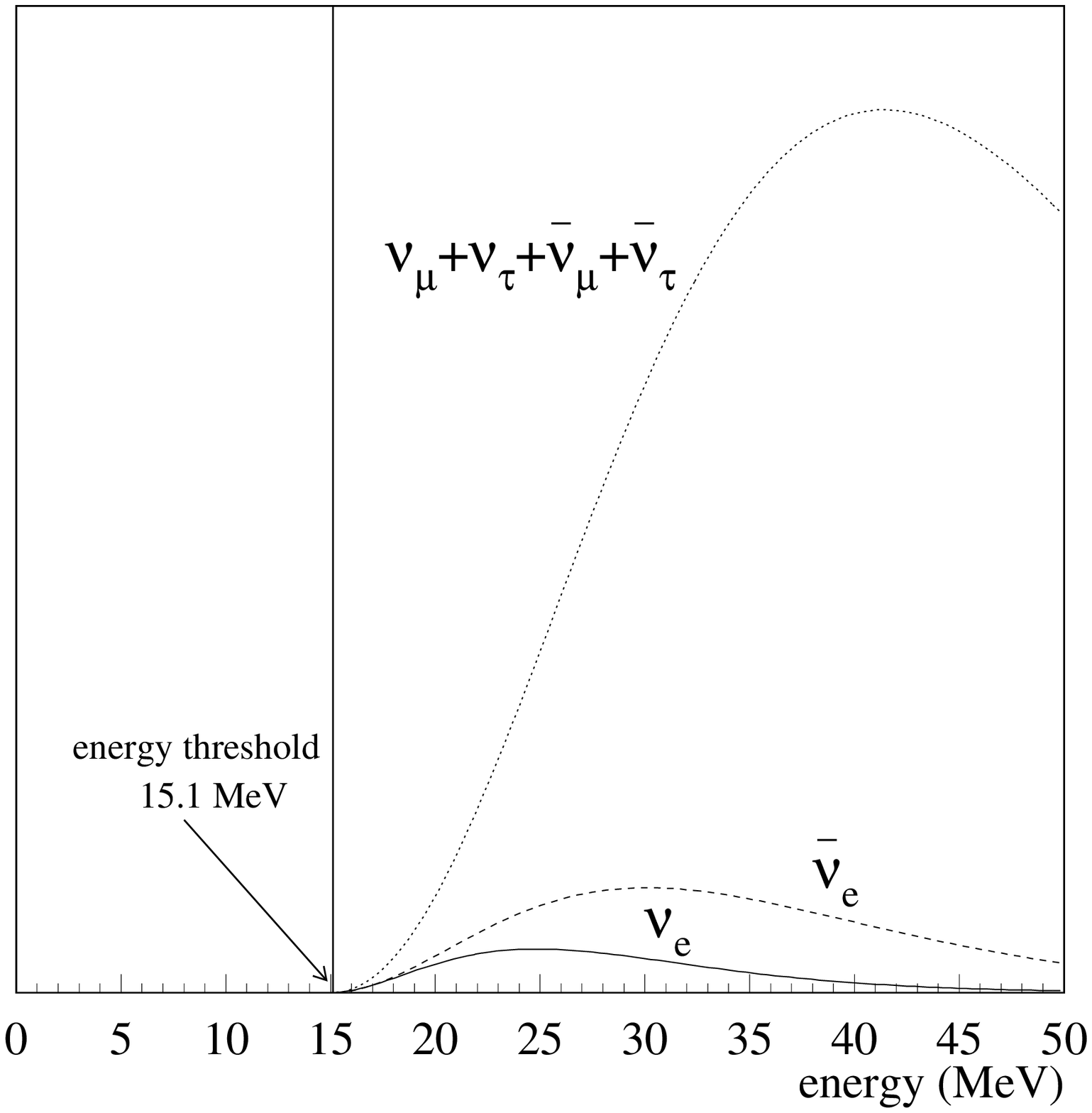} \\ [-0.4cm]
\mbox{\bf (a)}\;\;\; {\rm N(E_{\nu})} & 
\mbox{\bf (b)}\;\;\;{\rm N(E_{\nu})\,\sigma(E_{\nu})} \\
\end{array}$
\caption{Contribution of the different neutrino flavors to the 
neutral-current reaction 
$^{12}\mbox{C}(\nu,\nu')^{12}\mbox{C}^*(15.11\,{\rm MeV})$: 
{\bf (a)} supernova neutrino energy spectra and {\bf (b)}
their product with the cross section above threshold.
The solid line is the $\nu_e$ distribution; the dashed line is 
the $\bar{\nu}_e$ profile; and the dotted line is the summed 
contribution of the other flavors ($\nu_{\mu}$, $\nu_{\tau}$, 
$\bar{\nu}_{\mu}$ and $\bar{\nu}_{\tau}$).  
The reaction threshold is represented by the vertical line.}
\label{plotnc}
\end{figure}

In Borexino, the neutral-current excitation is dominated by
$\nu_{\mu}$, $\bar{\nu}_{\mu}$, $\nu_{\tau}$ and 
$\bar{\nu}_{\tau}$, due to their higher average energy;  
91\% of the neutral-current events come from the ``heavy flavor'' 
neutrinos.  
Their relative contribution to the neutral-current event rate is 
illustrated in Fig.~\ref{plotnc}.
The $\bar{\nu}_e - p$ charged-current events provide the time 
stamp for the ``massless'' species (possibly even the 
$\bar{\nu}_e - p$ events from SuperKamiokande).  
Thus, in Borexino, determining the time delay between the 
neutral-current and charged-current events provides a handle on 
the mass of $\nu_{\mu}$ and/or $\nu_{\tau}$.

Note that no energy information, E$_\nu$, is available from these 
neutral-current events.
Additionally, no distinction between $\nu_{\mu}$ and $\nu_{\tau}$ 
is possible. 
Nevertheless, the overall rate of these events constrains the 
average energy of these neutrinos, provided that the distance to 
the supernova and the overall luminosity can be determined by 
other means (and given the assumptions outlined earlier, 
such as equipartition of luminosity and a thermal distribution 
of energies).

Beacom and Vogel have shown~\cite{beacom1,beacom2} that 
model-specific details relating to the emission time profile of 
neutrinos from a supernova do not have a pronounced effect on the 
arrival time distribution.  They demonstrate that the 
overwhelming consideration in analyzing the time delay for 
massive neutrinos is the time constant of the exponential decay 
of the neutrino luminosity.  
They show, in addition, that averaged quantities such as:
\begin{equation}
\Delta t = \langle t \rangle_{NC}-\langle t\rangle_{CC}
\end{equation}
can be used as sensitive probes for extraction of the 
neutrino mass from supernova neutrino data.

\begin{figure} 
$\begin{array}{c c} 
\includegraphics[width=2.5in]{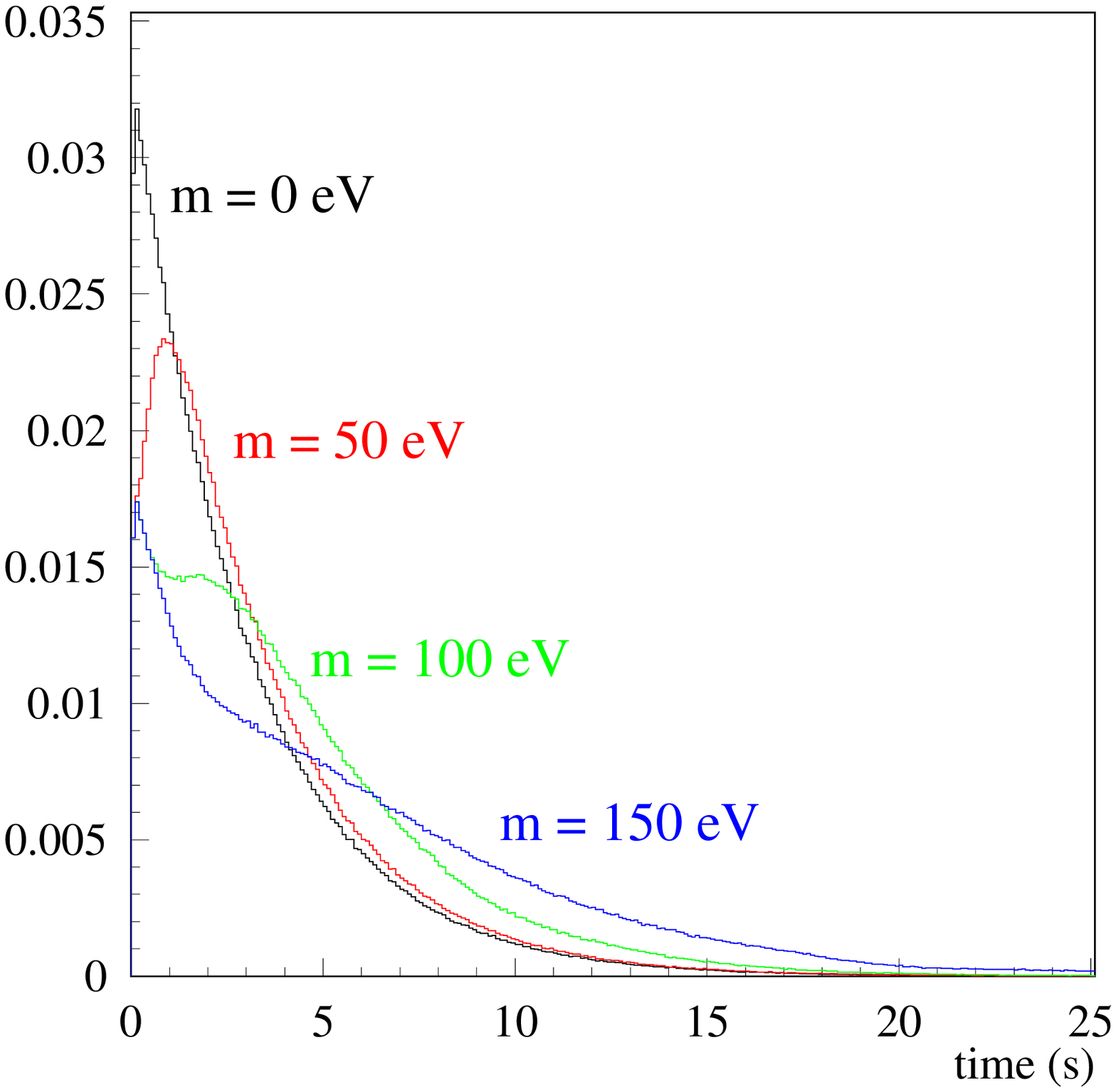} &
\includegraphics[width=2.5in]{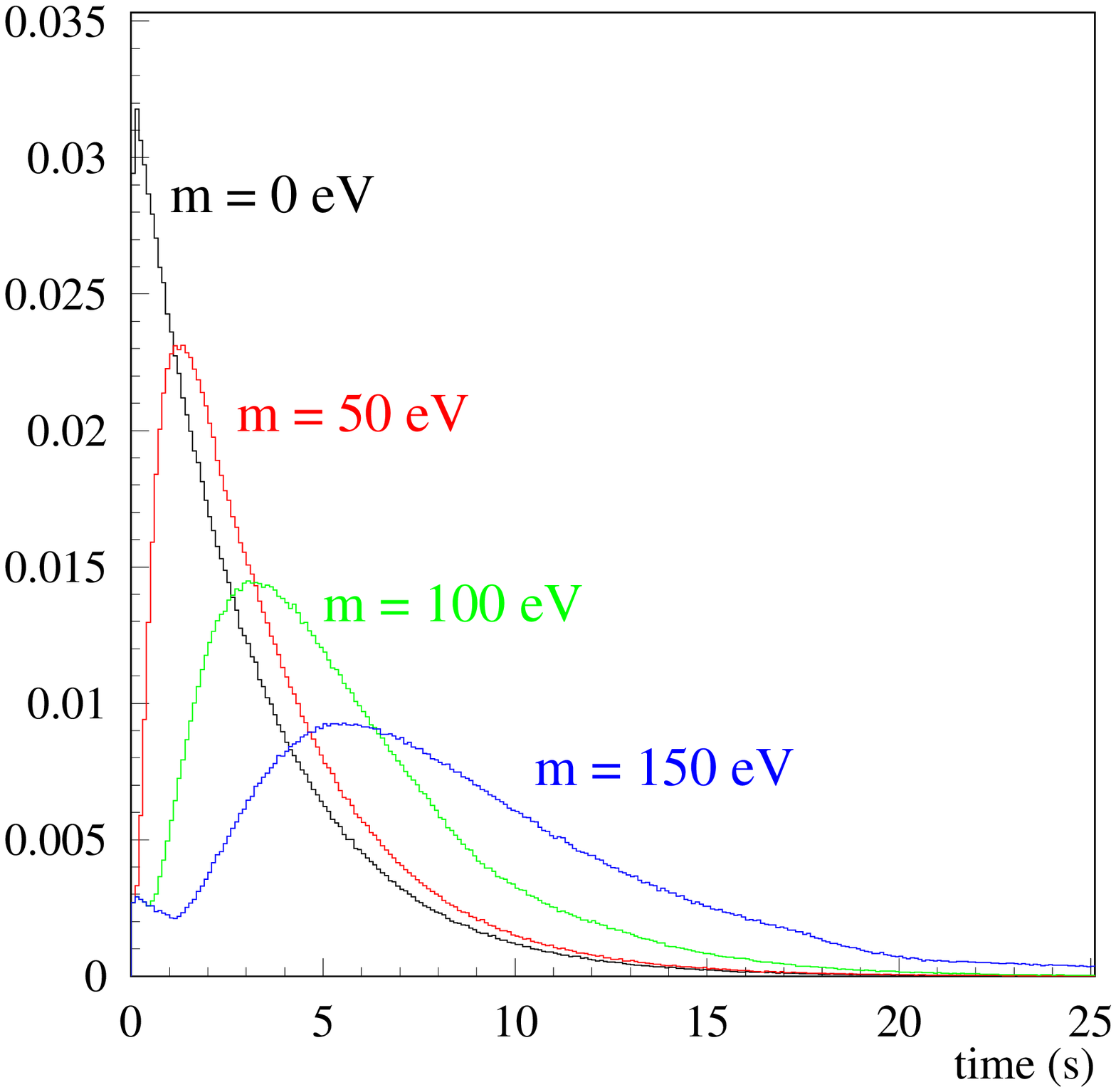} \\ [0.4cm]
\mbox{\bf (a)} \;\;\; {\rm only} \; {\rm m_{\nu_{\tau}}\neq 0} &
\mbox{\bf (b)} \;\;\; 
{\rm m_{\nu_{\mu}}\neq 0,\;m_{\nu_{\tau}}\neq 0} \\
\end{array}$
\caption{Time distribution for the 
$^{12}\mbox{C}(\nu,\nu')^{12}\mbox{C}^*$ events, for two cases:
{\bf (a)} $46\%$ of the events are from massive neutrinos
($\nu_{\tau}+\bar{\nu}_{\tau}$); 
{\bf (b)} $91\%$ of the events are massive
($\nu_{\mu}+\bar{\nu}_{\mu}$ and $\nu_{\tau}+\bar{\nu}_{\tau}$).}
\label{distributions}
\end{figure}

We considered a model for a supernova neutrino burst that rises 
linearly, reaching maximum in the first $20\,$ms.  
This is followed by an exponential decay with 
$\tau = 3\,$s~\cite{emtime}.
Fig.~\ref{distributions} shows the expected time distribution of 
the $^{12}\mbox{C}(\nu,\nu')^{12}\mbox{C}^*$ events in Borexino, 
for two scenarios. 
In both cases, $\nu_{e}$ is assumed massless and:
\begin{itemize}
\item[(a)] $\nu_{\mu}$ is massless and $\nu_{\tau}$ is massive
$\Rightarrow 46\%$ of the 
$^{12}\mbox{C}(\nu,\nu')^{12}\mbox{C}^*$ events are delayed; 
\item[(b)] $\nu_{\mu}$ and $\nu_{\tau}$ are both massive
$\Rightarrow 91\%$ of the 
$^{12}\mbox{C}(\nu,\nu')^{12}\mbox{C}^*$ events are delayed. 
\end{itemize}

\begin{figure} 
$\begin{array}{c c} 
\includegraphics[width=2.5in]{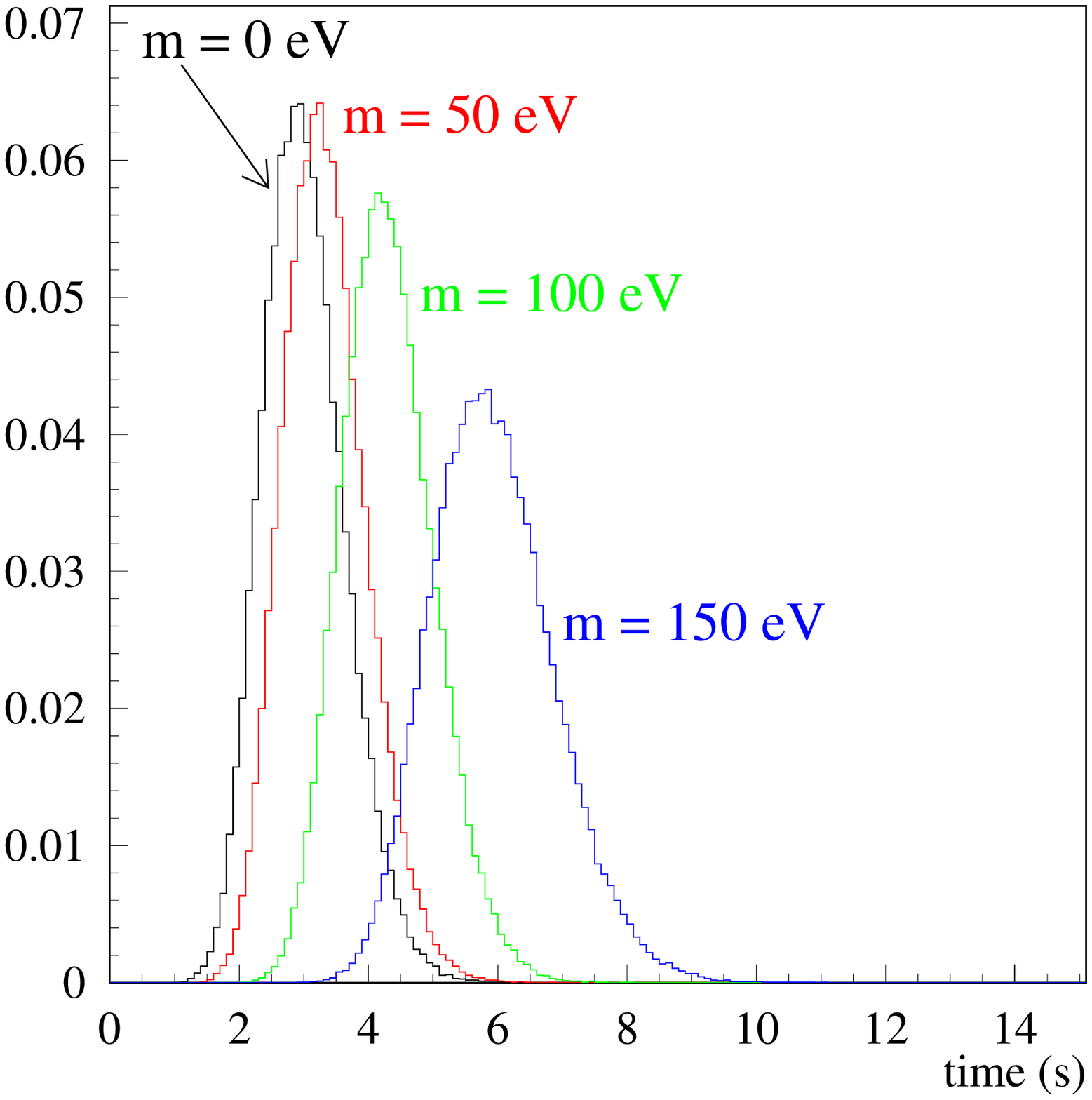} &
\includegraphics[width=2.5in]{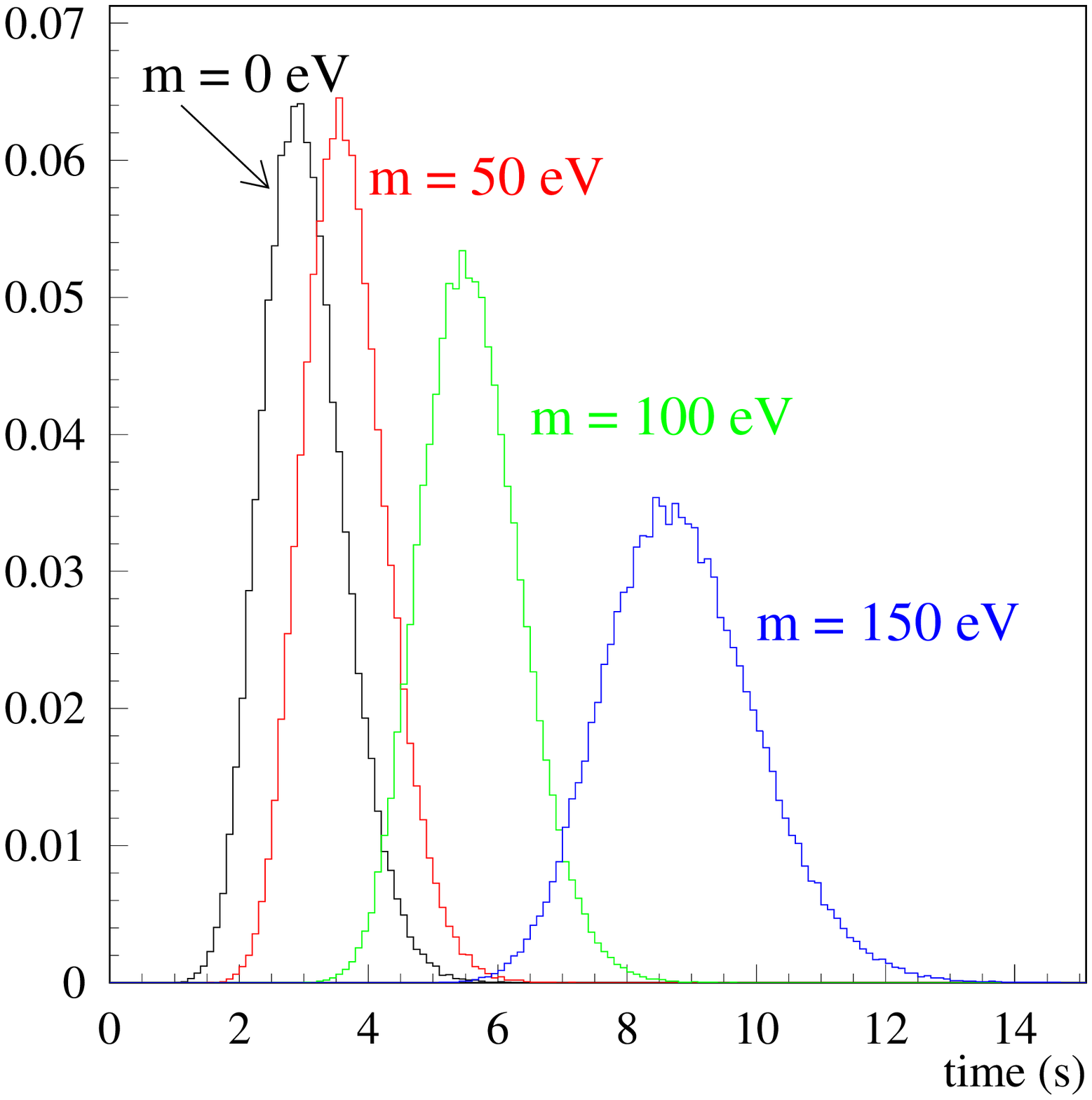} \\ [0.4cm]
\mbox{\bf (a)} \;\;\; {\rm only} \; {\rm m_{\nu_{\tau}}\neq 0} &
\mbox{\bf (b)} \;\;\; 
{\rm m_{\nu_{\mu}}\neq 0,\;m_{\nu_{\tau}}\neq 0} \\
\end{array}$
\caption{Monte Carlo distribution of the average time of the 
$^{12}\mbox{C}(\nu,\nu')^{12}\mbox{C}^*$ events, for a sample of
$10^5$ supernovae 
at $10\,$kpc (in this model, $t=3\,$s is the average arrival time
of the massless neutrinos).}
\label{averages}
\end{figure}

Fig.~\ref{averages} shows the distribution of average time of 
arrival of the $^{12}\mbox{C}(\nu,\nu')^{12}\mbox{C}^*$ events, 
for different values of the heavy neutrino mass. 
It is based on the results of a Monte Carlo simulation 
of $10^5$ supernovae (distance $10\,$kpc) producing $^{12}$C 
neutral-current events in Borexino.
The arrival times for the, on average, 23 events were drawn from 
the distributions shown in Fig.~\ref{distributions}.
Time zero is the theoretical instant of the earliest possible 
arrival (possibly determined by the earliest detection of 
$\bar{\nu}_e$ at SuperKamiokande, corrected for flight time).  
To obtain $\Delta t$, we subtract the average arrival time of 
the light neutrinos, that is $\langle t\rangle_{CC}=3\,$s.

\begin{figure}
$\begin{array}{c c} 
\includegraphics[width=2.5in]{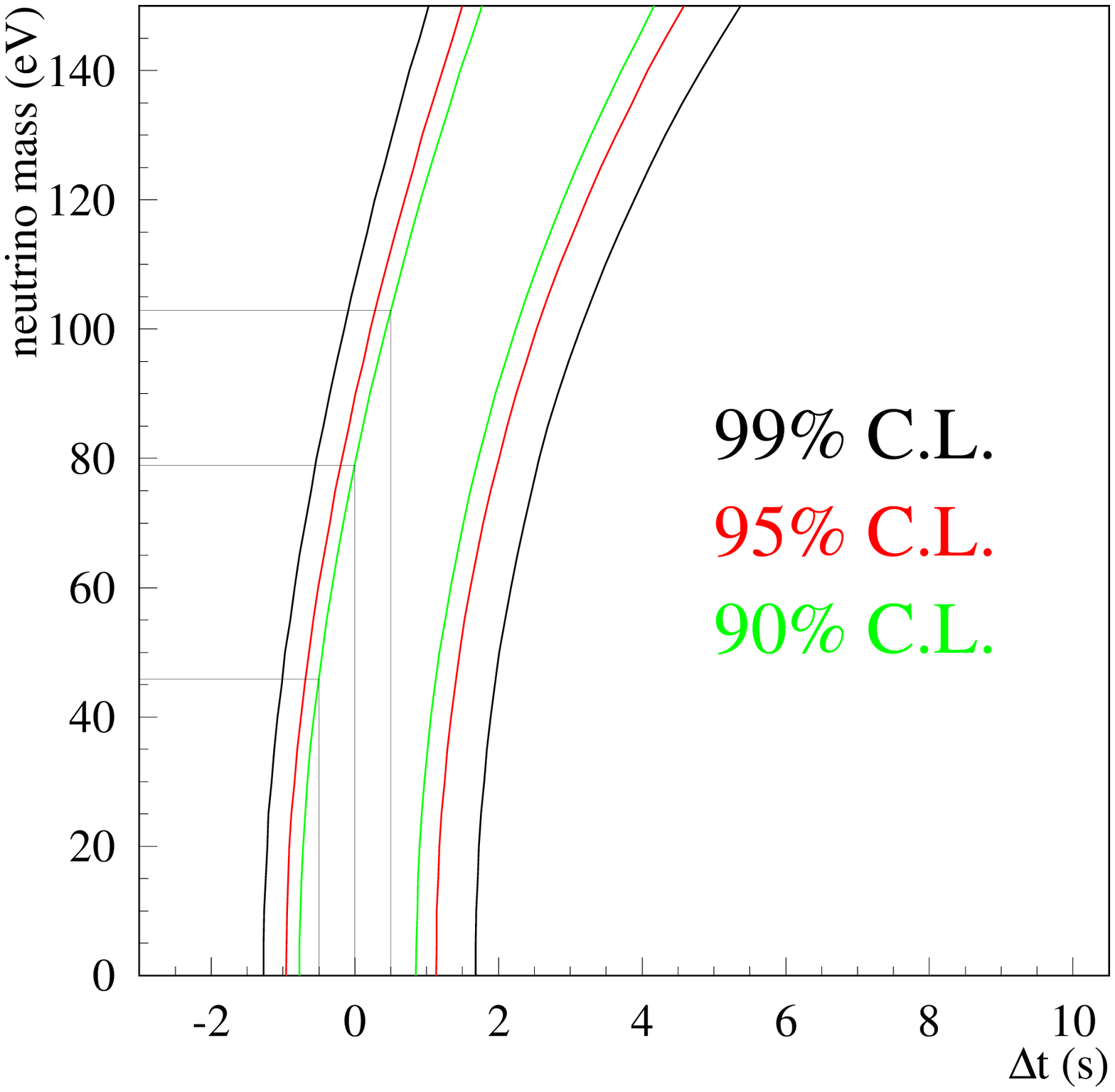} &
\includegraphics[width=2.5in]{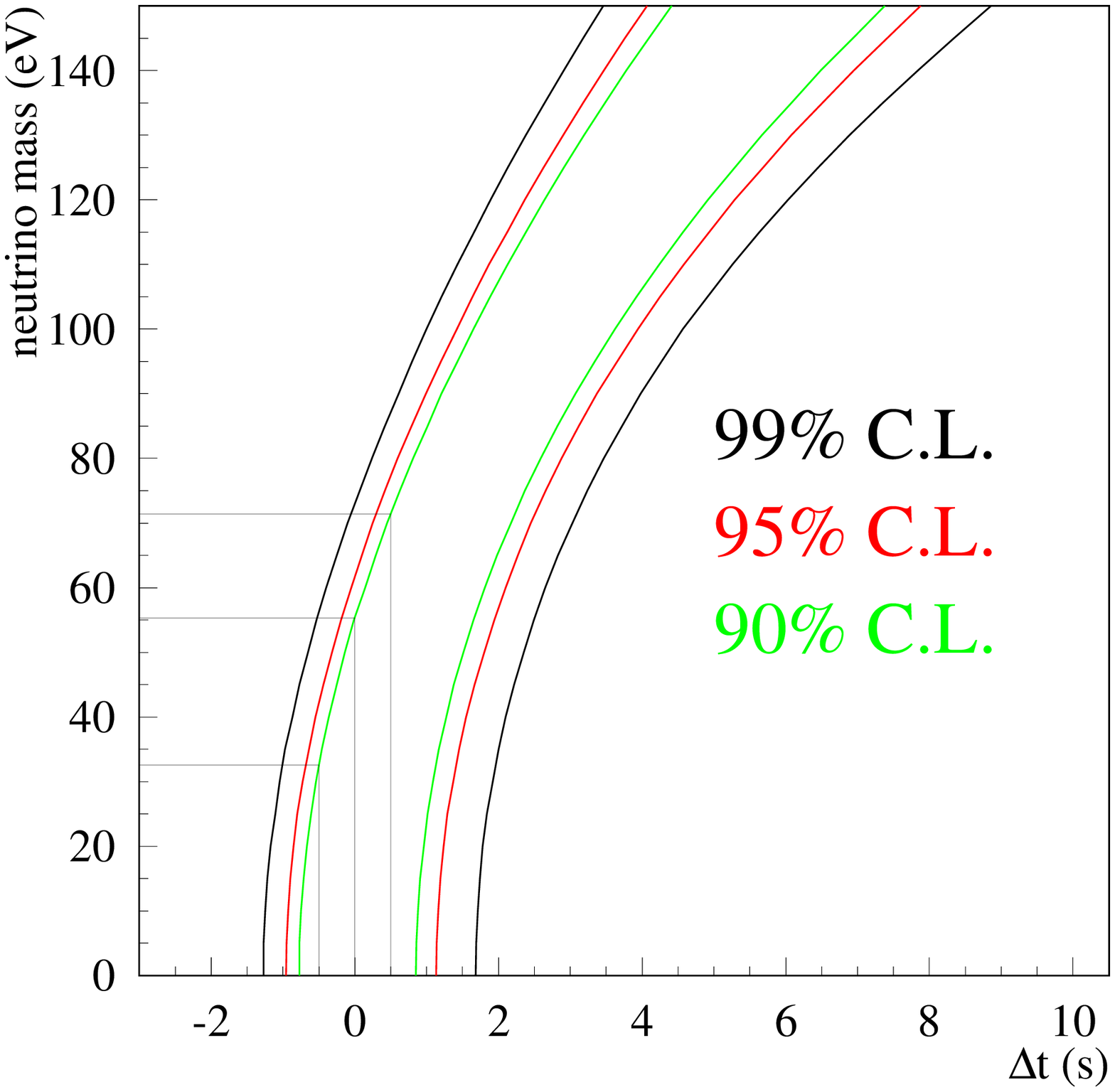} \\ [0.4cm]
\mbox{\bf (a)} \;\;\; {\rm only} \; {\rm m_{\nu_{\tau}}\neq 0} &
\mbox{\bf (b)} \;\;\; 
{\rm m_{\nu_{\mu}}\neq 0,\;m_{\nu_{\tau}}\neq 0} \\
\end{array}$
\caption{Contour probability plots relating the average delay
of $^{12}\mbox{C}(\nu,\nu')^{12}\mbox{C}^*$ events to the 
neutrino mass, in the two scenarios.
In both cases, the vertical lines correspond to a measured delay 
between charged- and neutral-current reactions equal to $-0.5$,
$0$ and $+0.5\,$s. Their intersections with the contour plots
set limits on the neutrino mass with 90\%, 95\% and 99\% C.L.
For instance,
if no delay is measured between the $p(\bar\nu_e,e^+)n$
charged-current events and the  $^{12}\mbox{C}(\nu,\nu')^{12}\mbox{C}^*$ 
neutral-current events, we can state with 90\% C.L. that 
$m_{\nu_{\tau}}< 79\,$eV in scenario (a) or 
$m_{\nu_{\mu\tau}}< 55\,$eV in scenario (b).}
\label{probs}
\end{figure}

Fig.~\ref{probs} shows the probability contour plots for neutrino 
mass as a function of average arrival time delay.
Different probability contours are included in the figure.
The interpretation of these plots, in terms of neutrino mass 
limits, goes as follows.  
Consider scenario (a), where $\nu_{\tau}$ is massive ($46\%$ of 
the neutral-current events are delayed).  From the data set,
the average arrival time of the charged-current events is 
subtracted from the average arrival time of the neutral-current 
events, giving a value for $\Delta t$.  
Given a measured delay (or lack thereof) allows one to set a 
mass limit, such that:
\begin{itemize}
\item[ ] if $\Delta t = -0.5\,$s, 
            then $m_{\nu_{\tau}}<46\,$eV $(90\%\; CL)$ 
\item[ ] if $\Delta t = 0\,$s,    
            then $m_{\nu_{\tau}}<79\,$eV $(90\%\; CL)$ 
\item[ ] if $\Delta t = +0.5\,$s, 
            then $m_{\nu_{\tau}}<103\,$eV $(90\%\; CL)$  
\end{itemize}

Similarly, for scenario (b) in which both 
$\nu_{\tau}$ and $\nu_{\mu}$ are massive, with 
$m_{\nu_{\mu}}\simeq m_{\nu_{\tau}}$ (as might arise for small
$\Delta m_{23}$) mass limits are extracted such that: 
\begin{itemize}
\item[ ] if $\Delta t = -0.5\,$s, then 
         $m_{\nu_{\mu\,\tau}}<33\,$eV $(90\%\; CL)$ 
\item[ ] if $\Delta t = 0\,$s, then 
         $m_{\nu_{\mu\,\tau}}<55\,$eV $(90\%\; CL)$ 
\item[ ] if $\Delta t = +0.5\,$s, then 
         $m_{\nu_{\tau}}<71\,$eV $(90\%\; CL)$ 
\end{itemize}

Conversely, a measured average arrival time delay of 
$\Delta t = +1.0\,$s allows one to exclude a massless 
$\nu_\tau$ at greater than 90\% confidence level.

These limits are only slightly worse than what might be 
achievable in SuperKamiokande 
(a sensitivity of $50\,$eV)~\cite{beacom1}, and 
in SNO (a sensitivity of $30\,$eV)~\cite{beacom2}. 
It is remarkable that the limits in Borexino are comparable, 
given the much higher statistics expected in these much larger 
detectors.
Since $\Delta t \propto {\rm m}^2$, and the error on $\Delta t$ 
is proportional to the square root of the number of
detected events, the error on the neutrino mass will, in the end,
be proportional only to the fourth root of the number of detected 
events.
This aspect combined with the higher ``heavy-flavor fraction''
in the neutral-current events in Borexino allow $\nu_\tau$ mass 
limits in the cosmologically-significant range to be reached.

These results, while being independent of many of the details in
the supernova neutrino emission models, are nevertheless based
on the assumptions made here on the supernova neutrino energy
spectrum and intensity.  In these calculations, the distance to
the supernova was assumed to be precisely determined.


\subsection{Neutrino Oscillations} \label{oscilla}
   
Neutrino oscillations can be probed by comparing the supernova 
neutrino event rates for different reactions.  
The extent of limits on ${\rm \Delta m^2}$ depend on the 
${\rm L/E}$ ratio which, for distances of kiloparsecs, is many 
orders of magnitude lower than present regions explored 
(e.g.\ solar neutrino vacuum oscillations).

We will consider, as an example, the implications of vacuum 
oscillations in the solar neutrino sector on the detection of 
supernova neutrinos in Borexino.  
Since the actual phase of the oscillation can be any value 
(varies rapidly over these distance scales) we take an average 
value of:
${\rm \langle  sin^2 ({\pi L}/\lambda_{osc})\rangle_L}=0.5$.

The main consideration is that higher energy $\nu_\mu$ could 
oscillate into $\nu_e$, resulting in an increased event rate 
since the expected $\nu_e$ energies are just at or below the 
charged-current reaction threshold.  The cross section for
$^{12}\mbox{C}(\nu_e,e^-)^{12}\mbox{N}$ increases by a
factor of 35 if we average it over a $\nu_e$ distribution with
$T=8\,$MeV, rather than $3.5\,$MeV. 
The gain in cross section for 
$^{12}\mbox{C}(\bar{\nu}_e,e^+)^{12}\mbox{B}$
is a factor of 5.  The large increase in the $\nu_e$ induced
reaction rate is a pseudo-appearance signature for oscillations.

\begin{figure}  
\includegraphics[width=5in]{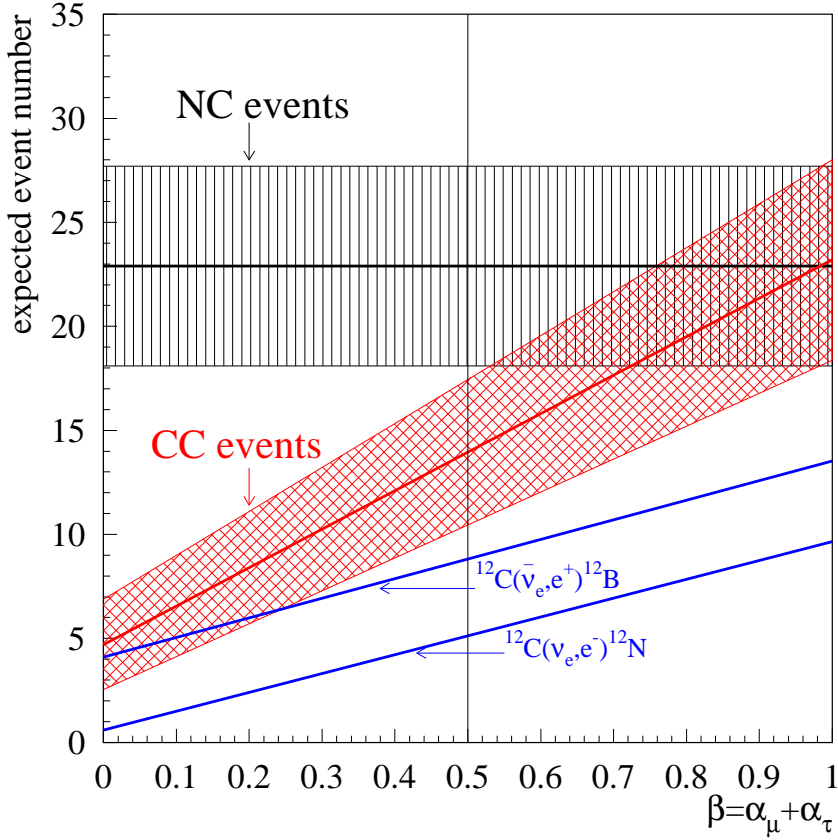}
\caption{Number of NC and CC events on $^{12}\mbox{C}$, as a 
function of an overall conversion probability 
$\beta=\alpha_{\mu}+\alpha_{\tau}$,
where $\alpha_{\mu}={\rm P}(\nu_e\leftrightarrow\nu_{\mu})$ and
$\alpha_{\tau}={\rm P}(\nu_e\leftrightarrow\nu_{\tau})$. The  
error bands refer to the case of a supernova at $10\,$kpc with
$\varepsilon_B=3\times 10^{53}\,$ergs.}
\label{oscfig}
\end{figure}

We assign the probability $\alpha_{\mu}$ to
$\nu_e \leftrightarrow \nu_{\mu}$ conversion and $\alpha_{\tau}$ 
to be the $\nu_e \leftrightarrow \nu_{\tau}$ conversion 
probability, and define $\beta=\alpha_{\mu}+\alpha_{\tau}$.  
Thus, depending on the value of $\beta$, we have the combinations
of event numbers reported in and displayed in Fig.~\ref{oscfig}; 
the number of charged-current and neutral-current events, as
a function of the parameter $\beta$ are shown with a 
$1\,\sigma$ error band corresponding to the statistics for a 
supernova at $10\,$kpc.  
The constant neutral-current rate fixes the flavor-independent 
luminosity.  
Though the statistics are low, the effect on the charged-current 
rates are significant. 
For instance, in the case of $\beta=0.5$, the number of
$^{12}\mbox{C}(\nu_e,e^-)^{12}\mbox{N}$ events increases 
from 0.65 to 5, while the number of 
$^{12}\mbox{C}(\bar{\nu}_e,e^+)^{12}\mbox{B}$ doubles, from 4 to 8.

\section{Conclusions}

The Borexino detector will be sensitive to neutrinos from a 
Type II supernova in our galaxy.  
A typical supernova at a distance of $10\,$kpc will produce
about 80 events from $\bar{\nu}_e$ capture on protons and about 
30 events from reactions on $^{12}\mbox{C}$, mostly from 
$\nu_\mu$ and $\nu_\tau$ neutral-current events.  
As a large homogeneous volume liquid scintillator, identification 
of the neutral-current events via detection of the monoenergetic 
$15.1\,$MeV $\gamma$ enables Borexino to explore non-standard
neutrino physics.  
Neutrino masses in the tens of eV range can be explored for 
$\nu_\mu$ and $\nu_\tau$ by measuring the delayed arrival of the 
neutral-current signal which is predominantly due to the 
interactions of $\nu_\mu$ and $\nu_\tau$.  
The charged-current and neutral-current reactions on 
$^{12}\mbox{C}$ also offer an important tool for probing
neutrino oscillations.  The distinctive charged-current reactions
feature delayed $\beta$ decays, correlated in time
and space with the neutrino interactions, allowing both $\nu_e$
and $\bar{\nu}_e$ to be easily identified in a
flavor-oscillation analysis.

At the time of submission of this manuscript, the authors
learned of the calculation by Beacom {\em et al.}~\cite{Beacomnew} 
that recoil protons from $\nu - p$ scattering might provide an 
abundant, detectable signal in Borexino.  
The general idea is that
though weak neutral-current cross sections are lower, the
neutrinos (and antineutrinos) from supernovae are abundant
in all three flavors, making the total flux more than
six times greater than the flux from just $\bar{\nu}_e$.
Furthermore, the $\nu_\mu$ and $\nu_\tau$ flavors are more
energetic, increasing the total event rate in this channel
such that it even exceeds the $\bar{\nu}_e - p$
charged-current event rate.  This provides Borexino with
several hundred supernova neutrino interactions, in the
event of a galactic supernova.  The challenge is detection
of the low-energy proton recoils.  As Borexino is designed
for a low energy threshold and since Borexino will exploit
pulse-shape discrimination in the liquid scintillator, it is
likely that the proton recoils will be cleanly detected.
What this enables in Borexino is sensitivity to the energy in
the neutral-current signal and much greater statistics.

\section{Acknowledgments}
The authors thank P. Vogel for early discussions and
J.F.~Beacom for interesting and valuable 
discussions of the arrival time profile for massive neutrinos 
from a supernova.
This work was supported in part by the National Science 
Foundation.


\end{document}